\documentclass[12pt,floats]{article}
\usepackage{amsfonts,color,epsfig}
 \usepackage{epstopdf}
\usepackage{graphicx}
\usepackage{multirow}
\usepackage[multiple]{footmisc}
\usepackage{booktabs}
\newcommand{\nc}{\newcommand}

\newcommand{\rnc}{\renewcommand}
\renewcommand{\thefootnote}{\fnsymbol{footnote}}
\makeatletter
\rnc{\theequation}{\thesection.\arabic{equation}}
\@addtoreset{equation}{section}
\makeatother
\nc{\fig}[5]{
\begin{figure}[!htbp]
    \begin{center}
    \leavevmode
    \centerline{
        \includegraphics[width=#1, height=#2]{#3}
        }
    \caption[]{#4}
    \label{#5}
    \end{center}
\end{figure}}
\nc{\figs}[8]{
\begin{figure}[!htbp]
    \begin{center}
    \leavevmode
    \centerline{
        \includegraphics[width=#1, height=#2]{#3}
        \includegraphics[width=#4, height=#5]{#6}
        }
    \caption[]{#7}
    \label{#8}
    \end{center}
\end{figure}}

\begin{document}
\begin{flushright}
{\today}\\
\end{flushright}
\vspace{5mm}
\begin{center}
{{{\Large {\bf
Lifshitz Topological Black Holes}}}}\\ [5mm]
{R.B. Mann\footnote{email: rbmann@sciborg.uwaterloo.ca; on leave from
Department of Physics and Astronomy, University of Waterloo,
Waterloo, Ontario, N2L 3G1, Canada} } \\[5mm]
{Kavli Institute for Theoretical Physics,
University of California, Santa Barbara, CA 93106, USA}
\end{center}
\begin{abstract}{\small
I find a class of black hole solutions to a (3+1)
dimensional theory gravity coupled to abelian gauge fields
with negative cosmological constant that  has been proposed as the dual theory to a Lifshitz theory describing critical
phenomena in (2+1) dimensions. These black holes are all asymptotic to  a Lifshitz fixed point geometry and depend on a single parameter that determines both their area (or size) and their charge.  Most of the solutions are obtained numerically, but an exact solution is also obtained for a particular value of this parameter.  The thermodynamic behaviour of large black holes is almost the same regardless of genus,
but differs considerably for small black holes.  Screening behaviour is exhibited in the dual theory for any genus, but the critical length at which it sets in is genus-dependent for small black holes. }
\end{abstract}


\vspace{0.3cm}

\hspace{11.5cm}{Typeset Using \LaTeX}
\newpage
\renewcommand{\thefootnote}{\arabic{footnote}}
\setcounter{footnote}{0}

\section{Introduction}

The scope of concepts and applications of holography has blossomed in recent years.  Realized in terms of  the AdS/CFT correspondence, holography asserts that gravitational dynamics in an asymptotically Anti de Sitter (AdS) spacetime can be mapped onto a (relativistic) conformal field theory (CFT) in one less dimension
\cite{AdSCFT}. This duality conjecture has proven to be a useful tool in understanding the behaviour of strongly interacting field theories, whose dual description is in terms of weakly coupled gravitational dynamics in a bulk spacetime that is asymptotically AdS.

Over the years holographic duality has been extended beyond high energy physics to a much broader class of spacetimes and dual physical systems.
Some investigations have extended these ideas to asymptotically de Sitter 
\cite{dSCFT} and asymptotically flat spacetimes \cite{MannMarolf}, where holographic renormalization has been shown to be a fruitful tool for understanding conserved quantities and gravitational thermodynamics. 
More recently holography has been qualitatively employed in improving our understanding of  transport properties of QCD quark-gluon plasmas 
\cite{AdSQG} and in the construction of
gravity models that are conjectured to be dual to various systems in atomic physics and condensed matter physics.  For example holography
 has been shown to be applicable to a class of strongly correlated 
electron and atomic systems that exhibit relativistic dispersion relations, and whose dynamics near  a critical point is well described by a relativistic CFT.   The AdS/CMat correspondence is being actively employed to study superconductivity \cite{AdSsuper}, the quantum Hall effect \cite{AdSqh}, and a number of other condensed matter
systems that can be described by CFTs \cite{Hartnoll}. 

A key feature of interest is the scaling property 
\begin{equation}\label{eq01}
t \to \lambda^z t  \qquad  \mathbf{x}  \to \lambda  \mathbf{x}
\end{equation}
exhibited by fixed points governing the behaviour of various condensed matter systems. This behaviour differs from that which arises in the conformal group
\begin{equation}\label{eq02}
t \to \lambda  t  \qquad  \mathbf{x}  \to \lambda  \mathbf{x}
\end{equation}
and is concretely realized in the (2+1) dimensional theory
\begin{equation}\label{eq03}
S = \int dt d^2\mathbf{x} \left( \dot{\phi}^2 - K (\mathbf{\nabla}\phi)^2 \right)
\end{equation}
known as the Lifshitz theory.   The action (\ref{eq03}) is invariant under the scaling
relation (\ref{eq01}) with $z=2$ and has
been used to model quantum critical behavior in strongly correlated electron systems \cite{Lifcorr,Vish}.

From a holographic perspective the conjectured dual to this system is a gravitational theory \cite{Kachru} whose equations of motion yielded solutions with spacetime metrics asymptotic to the form
\begin{equation}\label{eq04}
ds^2 = \ell^2 \left( -r^{2z}dt^2 +\frac{dr^2}{r^2}+ r^2 d\mathbf{x}^2 \right)
\end{equation}
where the coordinates $(t,r,x^1,x^2)$ are dimensionless.  These obey the scaling
relations 
\begin{equation}\label{eq05}
t \to \lambda^z t  \qquad  r \to \lambda^{-1}r \qquad \mathbf{x}  \to \lambda  \mathbf{x}
\end{equation}
and  the only length scale in the geometry is $\ell$.  Metrics asymptotic to (\ref{eq04})
can be generated as solutions to the equations of motion that follow from the action
\begin{equation}\label{eq06}
S = \int d^4x \sqrt{-g}\left(R-2\Lambda -\frac{1}{4}F_{\mu\nu}F^{\mu\nu}
-\frac{1}{12}H_{\mu\nu\tau} H^{\mu\nu\tau} - \frac{C}{\sqrt{-g}} \epsilon^{\mu\nu\alpha\beta}
B_{\mu\nu} F_{\alpha\beta}\right)
\end{equation}
where $\epsilon^{\mu\nu\alpha\beta}$ is the Levi-Civita tensor density and 
$F_{\mu\nu} = \partial_{[\mu}A_{\nu]}$ and $H_{\mu\nu\tau} = \partial_{[\mu}B_{\nu\tau]}$ are Abelian gauge fields that are topologically coupled with coupling constant $C$.
The quantity $\Lambda$ is the cosmological constant. In order to obtain the 
asymptotic behaviour (\ref{eq04}) these constants are given by  \cite{Kachru}
\begin{equation}\label{eq07}
\Lambda = - \frac{z^2+z+4}{2\ell^2}   \qquad   2z = (C\ell)^2
\end{equation}
in terms of the length scale $\ell$.

Recently spherically symmetric  solutions to the equations of motion that follow
from (\ref{eq06}) were obtained for the $z=2$. These solutions included a discrete
set of solutions known as Lifshitz stars as well as a continuous set of black hole solutions
having finite temperature \cite{DanThor}.   The purpose of this paper is to extend this work to the
case of topological black holes, namely those whose event horizons are toroidal or
of higher genus as a consequence of identifications made in the spacetime
\cite{toprefs}.  

These black holes differ considerably from one another depending on the size of
the event horizon, which is uniquely fixed in terms of their electric charge (or vice-versa).  Large black holes have thermodynamic behaviour that is essentially
the same regardless of the genus.  However small black holes have markedly different
genus-dependent thermodynamic behaviour.  The entropy of genus-1 black holes  
scales linearly with the temperature for all values of the black hole radius.  
Higher genus black holes must have a radius $r_h > 1/\sqrt{5}$, with units defined
as in eq. (\ref{eq04}).   As their radius
decreases, so does their charge, and as $r_h \to  1/\sqrt{5}$ these black holes
approach their extremal AdS counterparts \cite{negmass}
 and have the same asymptotic structure.   While most solutions must be obtained numerically, it is possible to find an exact black hole solution in the higher-genus case. This solution appears to correspond to its zero-mass AdS counterpart, though its asymptotic behaviour is given by (\ref{eq04}).   Although all these results are for $z=2$, generalization to higher $z$ should be  possible.

The solutions found here differ from those with obtained from gravity duals of other non-relativistic quantum systems \cite{othergal}, for which topological black hole solutions have also recently been found \cite{topgal}. 

\section{Lifshitz Asymptotics and Exact Solutions}

The field equations that follow from the action ({\ref{eq06}) are
\begin{eqnarray}
\nabla^{\nu}F_{\mu\nu} &=& -\frac{C}{6\sqrt{-g}}\epsilon_{\mu \nu \alpha \beta} H^{\nu \alpha \beta}
\label{eq08}\\
\nabla^{\tau}H_{\mu\nu\tau} &=& \frac{C}{2\sqrt{-g}}\epsilon_{\mu \nu \alpha \beta} F^{\alpha \beta}
\label{eq09}\\
G_{\mu\nu} - \frac{5}{\ell^2}g_{\mu\nu} &=& \frac{1}{2}\left(F_{\mu \tau}F_\nu^\tau -\frac{1}{4}g_{\mu\nu} F^2\right)+  \frac{1}{4}\left(H_{\mu  \sigma \tau}H_\nu^{\ \sigma \tau} -\frac{1}{6}g_{\mu\nu} H^2\right)
\label{eq10}
\end{eqnarray}
In order to solve these equations an ansatz is needed that preserves the basic symmetries under consideration. For the metric I shall take
\begin{equation}\label{eq11}
ds^2 = \ell^2 \left( -r^{4}f^2(r) dt^2 +\frac{g^2(r) dr^2}{r^2}+ r^2 d\Omega_k^2 \right)
\end{equation}
where
\begin{equation}\label{eq12}
d\Omega_k^2 = \left\{\begin{array}{ll} d\theta^2 + \sin^2\theta d\phi^2 &k=+1\\ d\theta^2 + 
\theta^2 d\phi^2
&k=0\\  d\theta^2 + \sinh^2\theta d\phi^2 &k=-1
\end{array} \right.
\end{equation}
is the metric for spatial sections at fixed $(t,r)$ corresponding to the genus $\textsf{g}=0$ (spherical), $\textsf{g}=1$ (flat/toroidal), and $\textsf{g}\geq 2$ (hyperbolic) cases.  Compact spatial sections at fixed $r$ are possible by making appropriate identifications
in the $\mathbf{x}$ coordinates \cite{wendy,outside}.

The gauge field strengths are
\begin{equation}\label{eq13}
F_{rt} = -2\ell g(r) h(r) f(r) r   \qquad  H_{r\theta\phi} = 2\ell^2 j(r) g(r) r  \left\{\begin{array}{ll} \sin\theta &k=+1\\ \theta  & k=0\\  \sinh\theta &k=-1
\end{array} \right.
\end{equation}
with all other components either vanishing or being given by antisymmetrization.

The field strength $F_{\mu\nu}$ is that of an electric field directed radially outward, and
described by the function $h(r)$ in an orthonormal basis.  This field sources the
3-form field strength $H_{\mu\nu\tau}$ and vice-versa.  This latter field is therefore electrically charged and  corresponds to a charged fluid whose density is given by $j(r)$
in an orthonormal basis \cite{DanThor}.

The field equations (\ref{eq08} -- \ref{eq10}) reduce to the system 
\begin{eqnarray}
r\frac{df}{dr} &=& -\frac{5}{2} f(r) + \frac{1}{2} f(r) g(r)^2 \left(5+\frac{k}{r^2} + j(r)^2-h(r)^2\right)
\label{eq14}\\
r\frac{dg}{dr} &=& \frac{3}{2} g(r) -  \frac{1}{2} g(r)^3 \left(5+\frac{k}{r^2} - j(r)^2-h(r)^2\right)
\label{eq15}\\
r\frac{dj}{dr} &=& 2 g(r) h(r) + \frac{1}{2}  j(r) - \frac{1}{2} j(r) g(r)^2 \left(5+\frac{k}{r^2} + j(r)^2-h(r)^2\right)
\label{eq16}\\
r\frac{dh}{dr} &=& 2 g(r) j(r) -2 h(r) 
\label{eq17}
\end{eqnarray}
which is a system of ODEs that can be solved by standard numerical methods.

An exact solution to the above equations  is
\begin{equation}
f(r) = j(r) = \sqrt{1+\frac{k \ell^2}{2r^2}}  \qquad  g(r) = \frac{1}{\sqrt{1+\frac{k \ell^2}{2r^2}}}
\qquad h(r) = 1
\label{eq18}
\end{equation}
yielding the metric
\begin{equation}
ds^2 = \left( -\frac{r^{2}}{\ell^2}\left(\frac{r^{2}}{\ell^2}+\frac{k}{2} \right)dt^2 +\frac{dr^2}{\left(\frac{r^{2}}{\ell^2}+\frac{k}{2} \right)}+ r^2 d\Omega_k^2 \right)
\label{eq19}
\end{equation}
where I have rescaled $r\to r/\ell$ and $t \to t/\ell$.
This solution is valid only for $z=2$ and does not straightforwardly generalize to other values of $z>1$.  

The spacetime described by (\ref{eq19}) is singular at $r=0$ for all values of $k$. For $k=0$ it 
becomes the metric (\ref{eq04}):  all its curvature scalars are finite but it is not geodesically complete .   This kind of metric can be regarded as physically reasonable if there exists a regular black hole solution that approaches it in some extremal limit \cite{Kachru,branefall}. For $k\neq 0$ the Kretschmann scalar diverges are $r=0$. If $k=1$ this is a naked singularity, but if $k=-1$ it is cloaked by an event horizon.   Hence the metric
\begin{equation}
ds^2 = \left( -\frac{r^{2}}{\ell^2}\left(\frac{r^{2}}{\ell^2}-\frac{1}{2} \right)dt^2 +\frac{dr^2}{\left(\frac{r^{2}}{\ell^2}-\frac{1}{2} \right)}+ r^2 (d\theta^2 + \sinh^2(\theta) d\phi^2) \right)
\label{eq20}
\end{equation}
is an exact black hole solution to the field equations provided $h(r)=1$ and  $j(r)= \sqrt{1-\frac{\ell^2}{2r^2}} $.  
The event horizon is located at $r=\ell/\sqrt{2}$. This metric would appear to be the analogue of a zero-mass topological AdS black hole \cite{negmass}, though it should be noted that defining mass and other conserved charges in this theory remains an open question at this stage.  

The asymptotic behaviour of the system (\ref{eq14})--(\ref{eq17}) is the same for all values of $k$, and so its analysis is identical to the given for the $k=1$ case \cite{DanThor}.  Equations  (\ref{eq15})--(\ref{eq17}) form a closed system for the set $\{g(r),j(r),h(r)\}$ and can be considered separately; once they are solved then eq. (\ref{eq14}) can be solved for $f(r)$.  Linearization of these equations indicates that there is a zero mode at large $r$.  However this mode must have zero amplitude if the system is to approach the Lifshitz metric (\ref{eq04}).   This 
can only happen if the initial values of the functions are appropriately adjusted (fine-tuned) to ensure that
each function in the set $\{g(r),j(r),h(r)\}$ approaches unity.  

\section{Series and Numerical Black Hole Solutions}

The system (\ref{eq14})--(\ref{eq17})  can be solved both for large $r$ and near the event horizon using series expansions.  For large $r$ the solution is
\begin{eqnarray}
f &=& 1+ {\frac {{\it k}}{4{r}^{2}}}- {\frac {{{\it k}}^{2}}{32{r}^{4}}}+ \left( {\frac {k^3}{128}} -\frac{k h_L}{12} \right) \frac{1}{r^6} + \left(\frac{13 k^2 h_L}{384} - \frac{5 k^4}{2048} 
+\frac{15  h^4_L}{64}\right) \frac{1}{r^8}
\nonumber\\
g &=&  1-{\frac {{\it k}}{4{r}^{2}}} + \left( \frac{h_L}{2}+{\frac {3k^2}{32}} \right) \frac{1}{r^4} - \left( {\frac {5k^3}{128}} + \frac{3k h_L}{8}  \right)  \frac{1}{r^6} + \left(\frac{7k^2 h_L}{32} + \frac{35 k^4}{2048} 
+\frac{3  h^4_L}{32}\right) \frac{1}{r^8}
\nonumber\\
j  &=&  1+{\frac {{\it k}}{4{r}^{2}}} - \left( \frac{3 h_L}{2}+{\frac {k^2}{32}}  \right) \frac{1}{r^4}
 + \left( {\frac {{k}^{3}}{128}} +{\frac {7  k h_L}{8}} \right)  \frac{1}{r^6} -
 \left(\frac{13 k^2 h_L}{32} + \frac{5 k^4}{2048}  +\frac{21  h^4_L}{32}\right) \frac{1}{r^8}
\nonumber\\
h &=&  1+ {\frac {h_L}{{r}^{4}}}- {\frac {k h_L}{2{r}^{6}}} + \left(\frac{7k^2}{32} h_L+ \frac{7}{16} h^2_L\right) \frac{1}{r^8}
\label{eq21}
\end{eqnarray}
and is governed by one constant $h_L$, that parametrizes how rapidly the electric field falls off at large $r$.
For $k=0$ the falloff is very rapid, particularly for $f(r)$, whose subleading term is proportional to $1/r^8$.

It is also possible to solve the equations near the event horizon.  Under the assumption that the black hole is not extremal, the metric functions $g_{tt}$ and $g_{rr}$ must respectively have a simple zero and a simple pole at the horizon $r=r_0$.  The electric field $h(r_0)=h_0$ is assumed to be finite at $r=r_0$.  Unlike the
situation for the (AdS)-Reissner-Nordstrom metric, these values are not independent because of the presence of the charged fluid (the 3-form field strength).  For a static configuration the gravitational pull of the black hole
must balance the self-repulsion of the fluid, whose value must also vanish at the horizon if the electric field is finite there.

Writing $r=r_0 + x$, the resultant near-horizon series solution is 
\begin{eqnarray}
f &=& f_0 \sqrt{x}\left(1 + f_1 x + f_2 x^2 +\cdots \right)
\nonumber\\
g &=&  \frac{g_0}{ \sqrt{x}}\left(1 + g_1 x + g_2 x^2 + \cdots \right)
\nonumber\\
j  &=&  j_0 \sqrt{x}\left(1 + j_1 x + j_2 x^2 + \cdots \right)
\nonumber\\
h &=&  h_0 \left(1 + h_1 x + h_2 x^2 + \cdots \right)
\label{eq21}
\end{eqnarray}
where 
\begin{eqnarray}
f_1 &=&  -\frac{5 k^2-52 r_0^4 h_0^2+45 k r_0^2-11 k r_0^2 h_0^2+6 r_0^4 h_0^4+100 r_0^4 }{2 (r_0 (k+5 r_0^2-r_0^2 h_0^2)^2}
\nonumber\\
g_0 &=&  \frac{ r^{3/2}_0}{\sqrt{(k+5 r_0^2-r_0^2 h_0^2)}}
\nonumber\\
g_1 &=&   \frac{3 k^2-24 r_0^4 h_0^2+25 k r_0^2-7 k r_0^2 h_0^2+4 r_0^4 h_0^4+50 r_0^4}{2 r_0 (k+5 r_0^2-r_0^2 h_0^2)^2}
\nonumber\\
j_0  &=& 2\frac{\sqrt{ r_0} h_0}{\sqrt{k+5 r_0^2-r_0^2 h_0^2}}
\nonumber\\
j_1  &=&  -\frac{ h_0 (k^2+11 k r_0^2-k r_0^2 h_0^2-8 r_0^4 h_0^2+30 r_0^4)}{2 r_0  (k+5 r_0^2-r_0^2 h_0^2)^2 }
\nonumber\\
h_1 &=& 2\frac{ (k+3 r_0^2-r_0^2 h_0^2)}{(k+5 r_0^2-r_0^2 h_0^2) r_0 }
\label{eq22}
\end{eqnarray}
with the remaining coefficients determined in terms of $h_0$ and $r_0$. 

From (\ref{eq22}) it clear that the constraint
\begin{equation}
(5-h^2_0) r^2_0 +k > 0
\label{eq23}
\end{equation}
must be respected so that the solutions are real.   For $k=0,1$ this imposes an upper bound on
$|h_0|$, whereas for $k=-1$ it imposes a lower bound of $r_0>1/\sqrt{5}$ on the size of the black hole.

Further progress can only be made numerically.  Since the equations are all first order ODEs, the shooting method can be used to solve them.  Choosing some value of $r_0$ for the black hole, one can then choose a value for $h_0$ and then find the initial values of
the functions $g(r)$, $h(r)$, and $j(r)$ from the series solutions for some value of $r>r_0$.
This becomes the initial data for the system  (\ref{eq15})--(\ref{eq17}), which can then be numerically solved. The value of $h(r)$ can be computed for $r=R>>r_0$; should 
$|h(R)-1|>\varepsilon$, where $\epsilon<<1$ is some level of tolerance, then the process is repeated, with value of $h_0$  adjusted, until $|h(R)-1| < \varepsilon$.   This then gives  a black hole solution of radius $r_0$ with asymptotic behaviour given by the metric
(\ref{eq04}).

In practice it proves best to begin with large black holes, for which the value of $h_0$ remains constant.  Numerically I find that setting $h_0=1.374$ yields valid solutions for $r_0>10$ for all values of $k$.  For $k=0$ this value of $h_0$ yields black hole solutions for
all values of $r_0$ that were computationally viable.  For $r_0<10$ it is necessary to systematically adjust the value of $h_0$ upward for $k=1$ and downward for $k=-1$.

Figure \ref{f1} illustrates the behaviour of $h_0$ as a function of $r_0$ on a log-log plot. 
For $k=1$, the value of $h_0$ asymptotes to the upper bound given in (\ref{eq23}).  The 
extremal limit appears to occur when the bound is saturated as $r_0 \to 0$; it does not appear to be possible to find series solutions for $r_0 >0$ in the extremal case.
For $k=-1$ the value of $h_0$ approaches zero as $r_0\to 1/\sqrt{5}$.  In this case the limit is evidently that of an extremal topological AdS black hole with genus $\textsf{g}>2$.  
 \begin{figure}
	\includegraphics[scale=0.15]{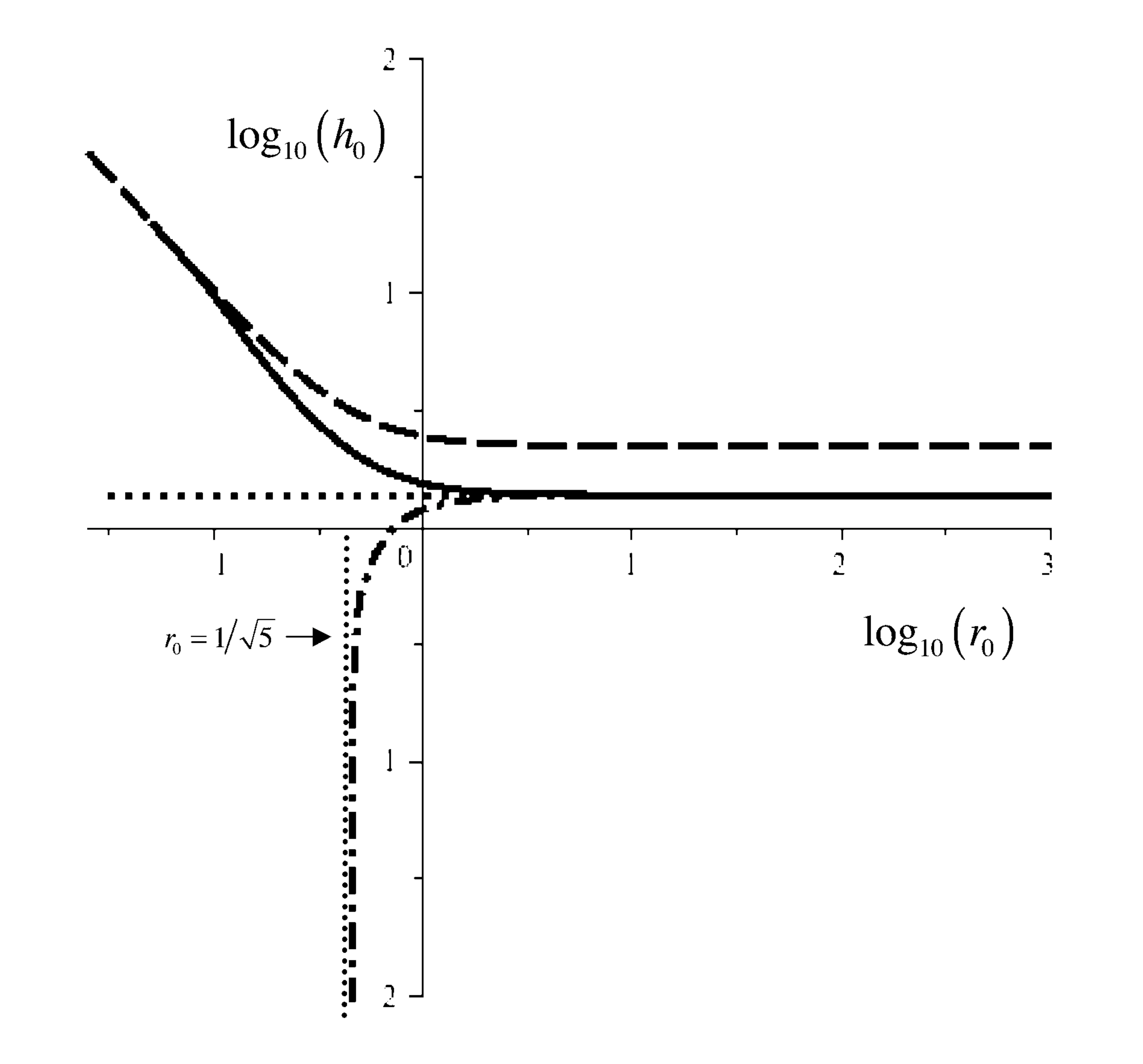}
	\caption{$h_0$ vs. $r_0$ on a log-log scale for spherical ($k=1$, solid), toroidal 
	($k=0$, dot) and higher-genus ($k=-1$, dot-dash) black holes.  The limit given
	in eq. (\ref{eq23}) for $k=1$ is the dashed line. The intersection point of
	the $k=-1$ curve with the $r_0$ axis corresponds to the exact solution
	(\ref{eq20}).}
	\label{f1}
	\end{figure}
For the exact solution (\ref{eq20}), $r_0=1/\sqrt{2}$ and $h_0=h(r)=1$, corresponding
to the intersection point of the $k=-1$ curve with the $r_0$ in figure \ref{f1}.   The exact solution provides a useful check on the numerical solution, and it can be verified that
the  solutions for the metric and gauge functions are identical to within limits of tolerance
(taken to be $10^{-4}$).

The metric functions and gauge field strengths are essentially indistinguishable amongst the different values of $k$ for 
$r_0>10$, as an example illustrates in figure \ref{f2}.  Small differences begin to appear for intermediate values of $r_0$, as shown in figure \ref{f3}, and for small $r_0$ the distinctions are quite significant, as depicted in figures \ref{f4} and \ref{f5}.
\begin{figure}
	\includegraphics[scale=0.15]{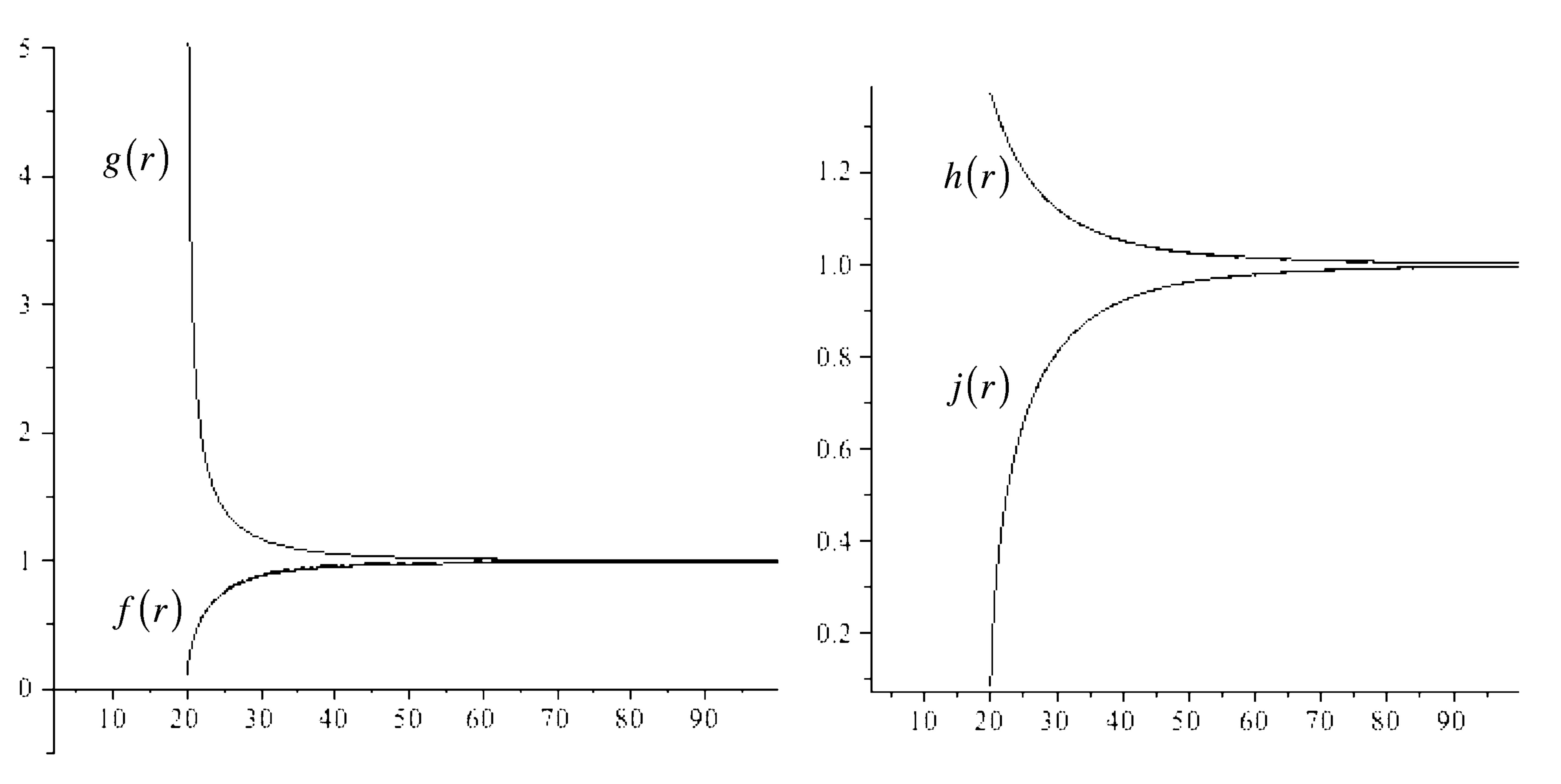}
	\caption{A plot of the metric and gauge functions for $r_0=20$ for all three values of
	$k$. The three curves overlap within the plotting resolution.}
	\label{f2}
\end{figure}
\begin{figure}
	\includegraphics[scale=0.15]{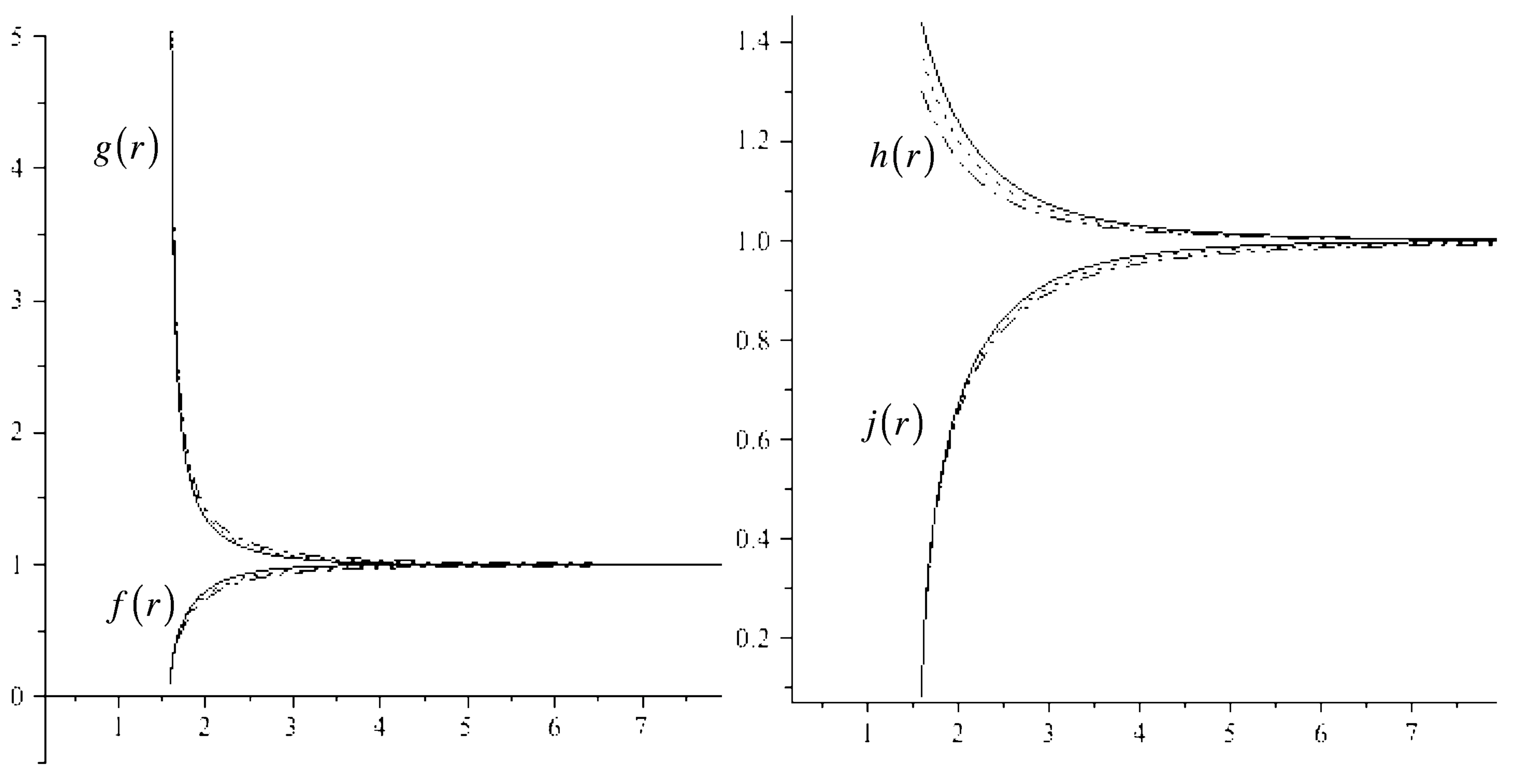}
	\caption{A plot of the metric and gauge functions for $r_0=1.6$ for 
	$k=1$ (solid), $k=0$ (dot), and $k=-1$ (dot-dash). }
	\label{f3}
\end{figure}
\begin{figure}
	\includegraphics[scale=0.15]{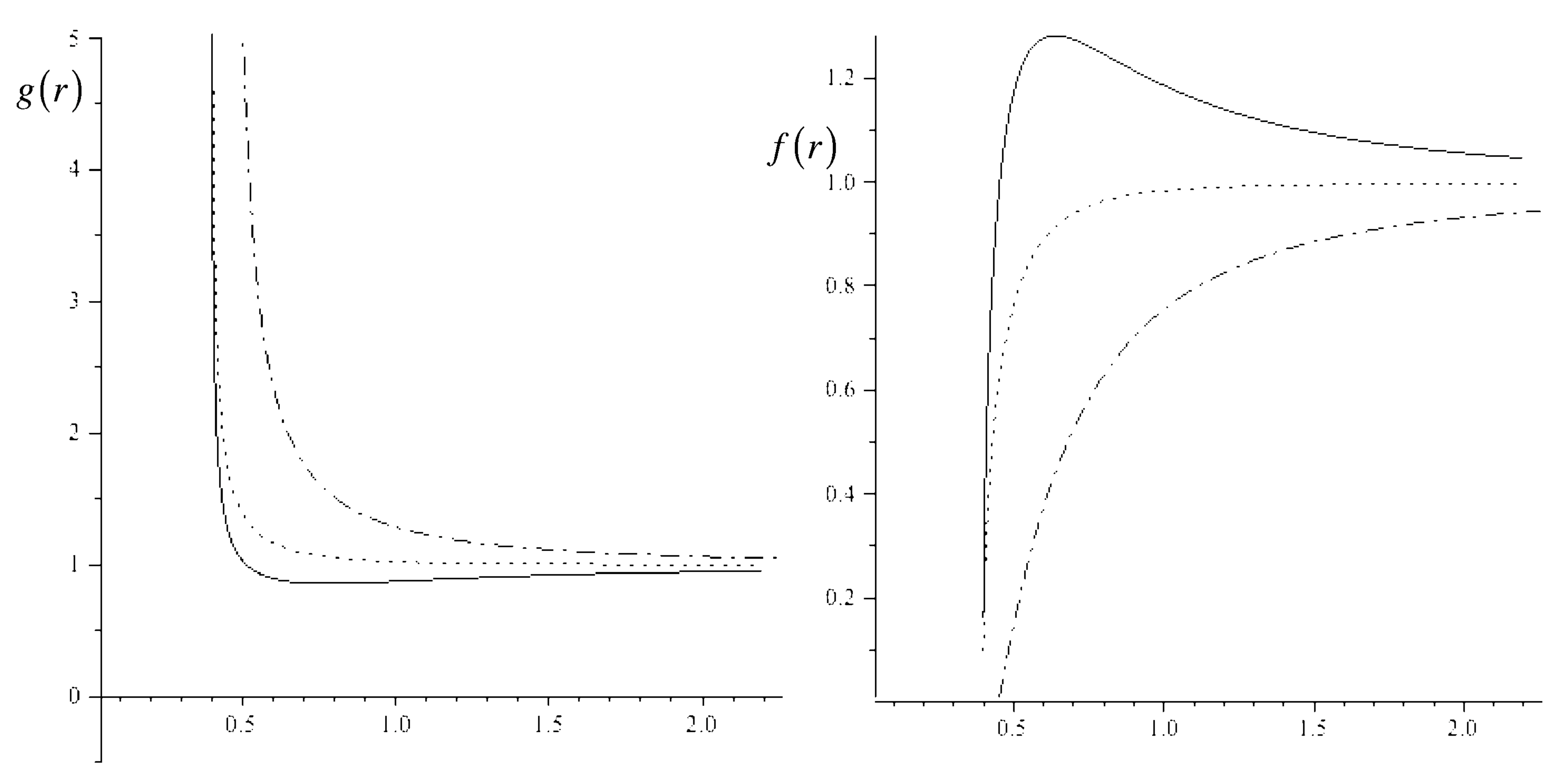}
	\caption{A plot of the metric functions for $r_0=0.4$ for 
	$k=1$ (solid), $k=0$ (dot), and $k=-1$ (dot-dash).}
	\label{f4}
\end{figure}
\begin{figure}
	\includegraphics[scale=0.15]{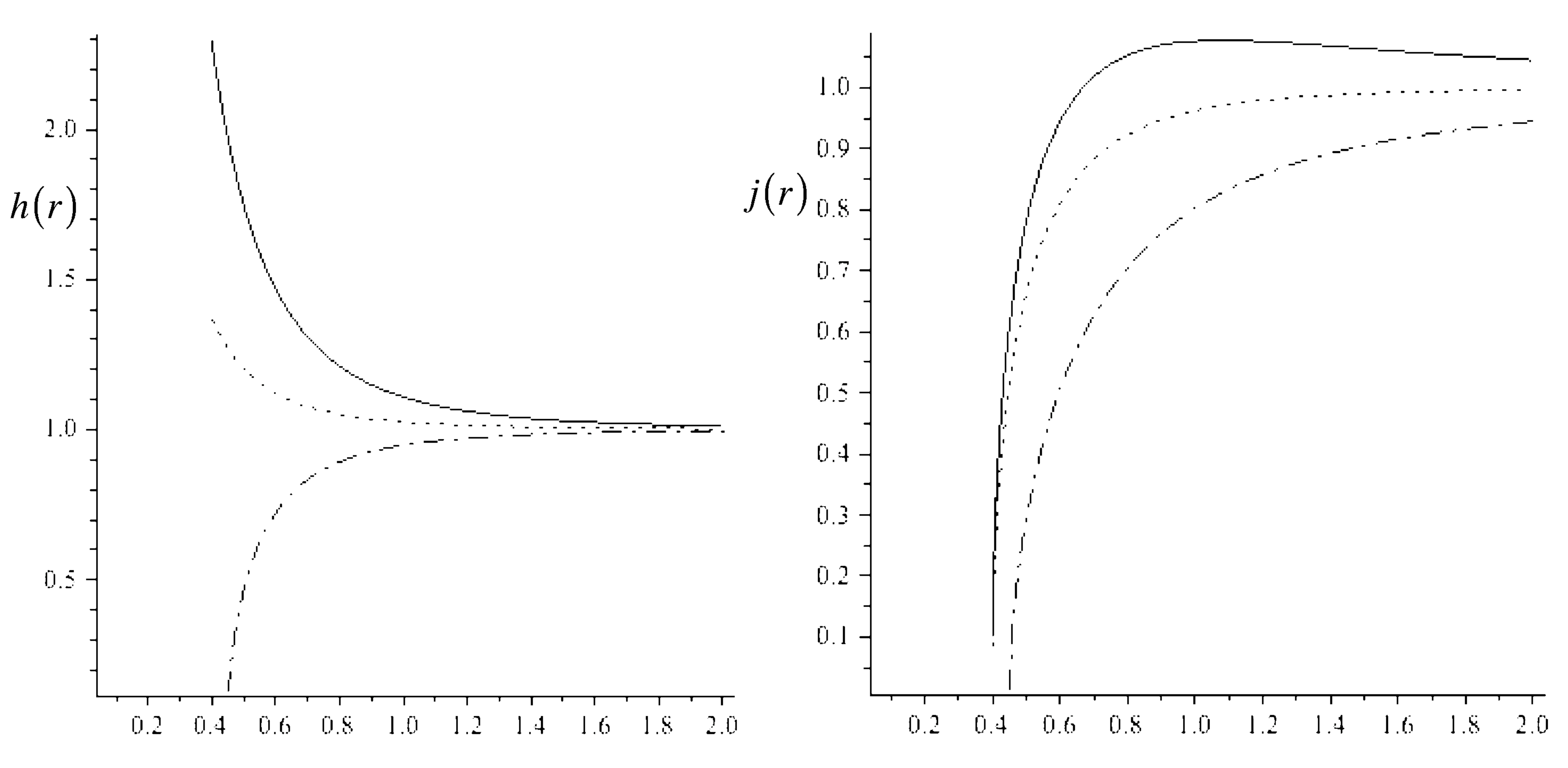}
	\caption{A plot of the gauge functions for $r_0=0.4$ for 
	$k=1$ (solid), $k=0$ (dot), and $k=-1$ (dot-dash).}
	\label{f5}
\end{figure}

\section{Black Hole Thermodynamics}

The temperature of Lifshitz black holes is easily evaluated using standard Wick-rotation methods, yielding the result
\begin{equation}
T = \frac{f_0 r_0^3}{4\pi g_0}=\frac{f_0 r_0^{3/2} \sqrt{-h_0^2 r_0^2+k+5 r_0^2} }{4 \pi}
\label{eq24}
\end{equation}
where $f_0$ is dependent upon $r_0$ so that the metric has the asymptotic behaviour given in equation (\ref{eq04}).  
\begin{figure}
	\includegraphics[scale=0.15]{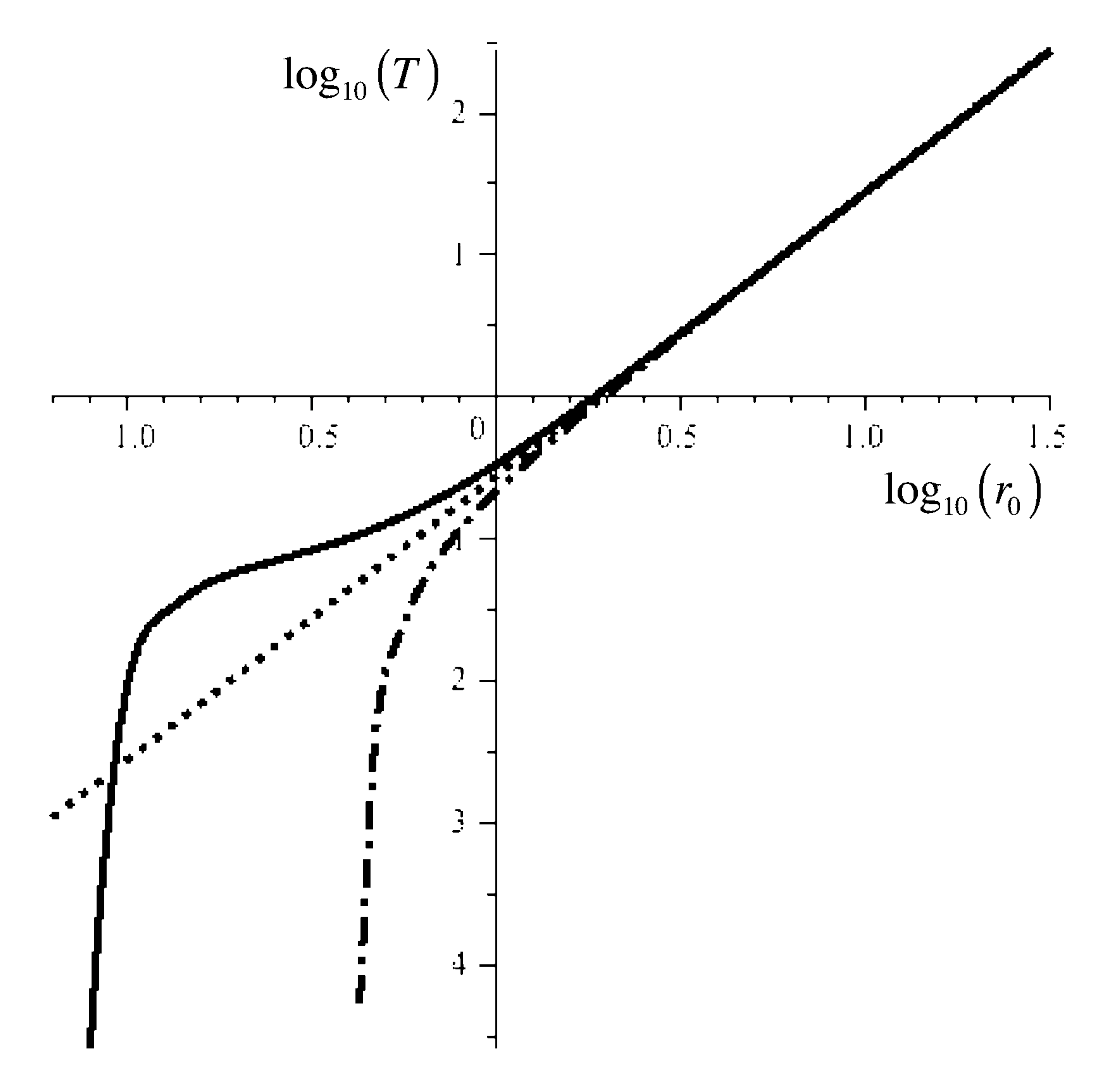}
	\caption{Dependence of the black hole temperature as a function
	of $r_0$ for $k=1$ (solid), $k=0$ (dot), and $k=-1$ (dot-dash).}
	\label{f6}
\end{figure}

The temperature for any genus is therefore dependent only on the size $r_0$ of the black hole, and its behaviour is illustrated in figure \ref{f6}.
For large $r_0$, spherical black holes ($k=1$) are hotter than 
planar/toroidal black holes ($k=0$), which are in turn hotter than the
$k=-1$ higher genus class.  The planar/toroidal holes have a temperature
dependence that increases quadratically with the black hole radius; numerically I find that
\begin{equation}
T_{k=0} = 10^{-0.57} r^2_0 = .276 r^2_0
\label{eq25}
\end{equation}
a behaviour that also accurately describes the $k=\pm 1$ cases for
$r_0 > 0.6$. This behaviour is consistent with the near-horizon expansion that indicates  $g_0\sim 1/f_0 \sim \sqrt{r_0}$ for large $r_0$.
\begin{figure}
	\includegraphics[scale=0.15]{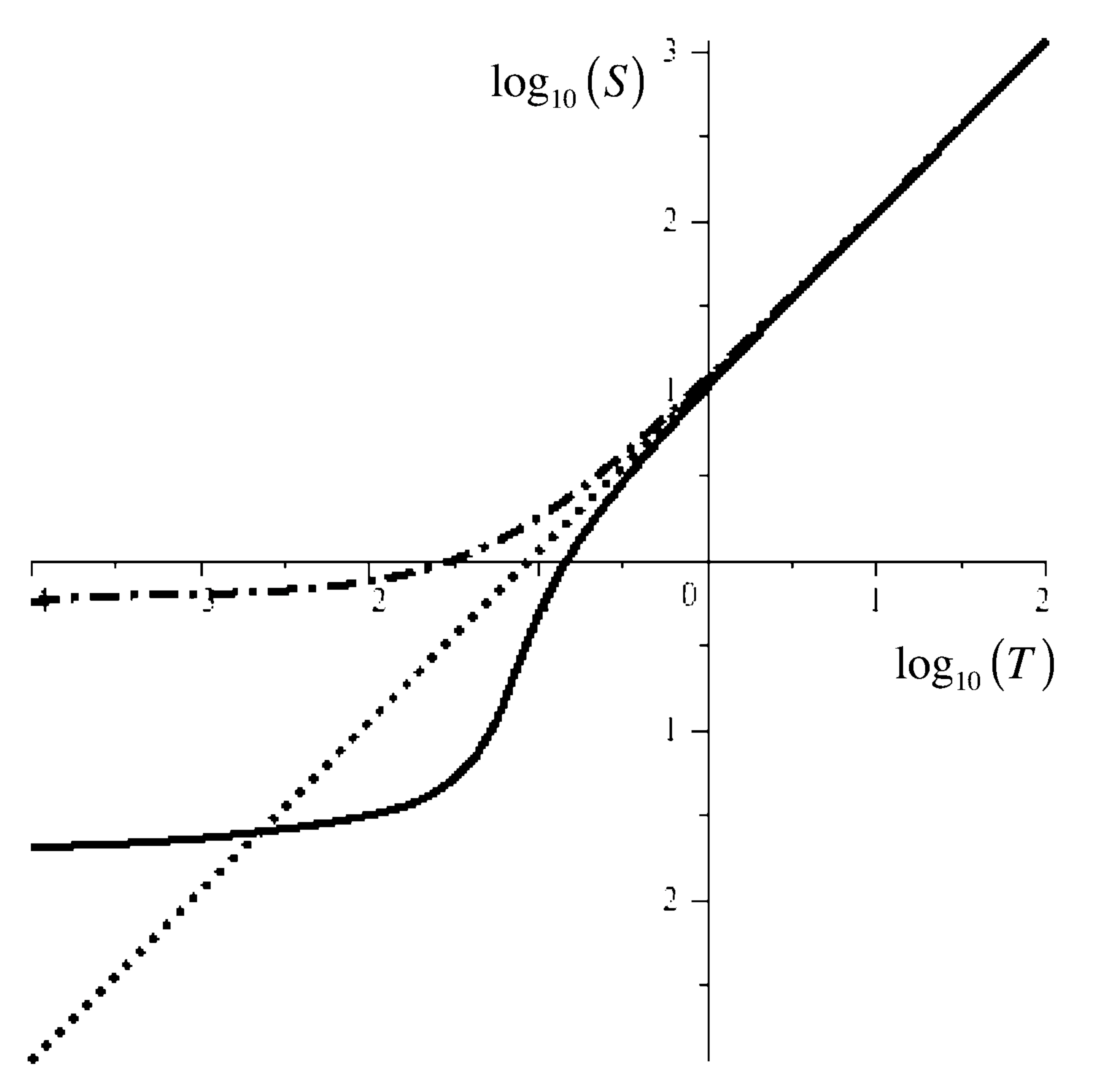}
	\caption{Dependence of the black hole entropy as a function
	of temperature for $k=1$ (solid), $k=0$ (dot), and $k=-1$ (dot-dash).}
	\label{f7}
\end{figure}
 
For small $r_0$ the behaviour is strikingly different for each case.  
Higher genus black holes approach zero temperature at an exponentially
rapid rate as $r_0 \to 1/\sqrt{5}$.   For $r_0 < 0.1$ the planar black holes
become hotter than their $k=0$ counterparts.  The temperature of the
$k=1$ black holes falls off very rapidly as $r_0$ decreases, though less so than the $k=-1$ case.  

The entropy of the black holes is given by $\frac{1}{4}$ of the area, so
$S = \pi r^2_0$ for all cases (assuming appropriate identifications and
volume normalizations for $k=0,-1$).  Hence   eq. (\ref{eq25}) gives
\begin{equation}
S = 11.4 T 
\label{eq26}
\end{equation}
a relationship that is valid for all $T$ for the $k=0$ case  and that
holds for large $T$ for $k=\pm 1$ \cite{DanThor}.  The behaviour of the entropy as a function of temperature is given in figure \ref{f7}.

For the exact solution (\ref{eq20}), the temperature and entropy are
\begin{equation}
 T = \frac{1}{4\pi}   \qquad   S = \frac{\pi}{2} 
\label{eq27}
\end{equation}
which can be computed directly from the exact solution or from the series solution (\ref{eq22}) for which $f_0=1/g_0= 2^{3/4}$ and
$r_0=1/\sqrt{2}$.  These values are commensurate with those in figure 
\ref{f7}, though they are considerably beyond the range for which
eq. (\ref{eq25}) applies.

\section{Wilson Loops and the Boundary Dual Theory}

Since in (2+1) dimensions one can write $\nabla^2\phi = \vec{\nabla}\times\vec{E}$, where
$E_j = \varepsilon_{jk}\nabla^k\phi$, the boundary theory (\ref{eq03}) can be regarded 
as a gauge theory in (2+1) dimensions (albeit one with an unusual action) with a dimensionless coupling
constant \cite{Vish,DanThor}. As in the $k=0$ spherical case, one can introduce Wilson loops by joining 
charged particles on the boundary that are connected together in the bulk via a string.  The 
Euclidean action
of this string for a rectangular Wilson loop is the same for all values of $k$ and is given by \cite{DanThor,Wilsloop}
\begin{equation}
{\cal S} = \frac{1}{2\pi\alpha^\prime}\int d\sigma d\tau \sqrt{\det[g_{AB}\partial_\mu X^A \partial_\nu X^B]}
=  \frac{\Delta \ell^2}{2\pi\alpha^\prime}\int d\theta \sqrt{f^2 r^{2z+2}+f^2 g^2 r^{2z-2}\left(\frac{dr}{d\theta}\right)^2}
\label{eq28}
\end{equation}
taking $\sigma=\theta$ and $i\tau=t$ in the static gauge, with Euclidean time interval $\Delta$.  

Extremizing the action yields a constant of the motion
\begin{equation}
 \frac{f^2 r^{2z+2}}{ \sqrt{f^2 r^{2z+2}+f^2 g^2 r^{2z-2}\left(\frac{dr}{d\theta}\right)^2}} = f(r_m) r_m^{z+1}
\label{eq29}
\end{equation}
from which can be computed the boundary length
\begin{equation}
L = \int d\theta = 2 \int_{r_m}^\infty \frac{dr}{r^2}\frac{g}{\sqrt{\left(\frac{f}{f_m}\right)^2 \left(\frac{r}{r_m}\right)^{2z+2} -1 }}
\label{eq28}
\end{equation}
and the regularized potential energy between the two particles
\begin{equation}
V  = \frac{{\cal S}}{\Delta \ell} =  \frac{\ell}{2\pi \alpha^\prime}\left( 2 \int_{r_m}^\infty dr \frac{r^{z-1} f g}{\sqrt{1-\left(\frac{f_m}{f}\right)^2 \left( \frac{r_m}{r}\right)^{2z+2} }} - 2 \int_{r_0}^\infty dr r^{z-1} f g
  \right)
\label{eq29}
\end{equation}
where $f_m=f(r_m)$ and $r_m > r_0$ is the location of the midpoint of the string.  

For small $r_m$ the separation $L$ between the particles grows, reaching a maximum and then decreasing as $r_m$ gets larger.   The energy between the particles will is negative for sufficiently small $r_m$, and will
vanish at some particular value of $r_m$ (or $L$).    Beyond this point the energy for a single string joining
the pair is positive.  The energy of the configuration is therefore minimized (to zero) by 
two (non-interacting) strings stretching from each particle down to the horizon, screening the gauge
interaction between the particles.  The point at which this takes place will depend on $r_0$, and hence the temperature.  

For large black holes the metric functions $f$ and $g$ are nearly indistinguishable for all values of $k$,
and so the analysis of screening behaviour for the spherical $k=0$ case 
\cite{DanThor} holds for the other values of
$k$ as well.  For small black holes the value of $r_m$ at which the potential vanishes is numerically almost the same, as in figure \ref{f8}.  However the length between the particles differs considerably as a function of $r_m$ (see figure \ref{f9}), and so the critical value $L=L_c$ at which screening occurs will differ considerably for different values of $k$. 
\begin{figure}
	\includegraphics[scale=0.15]{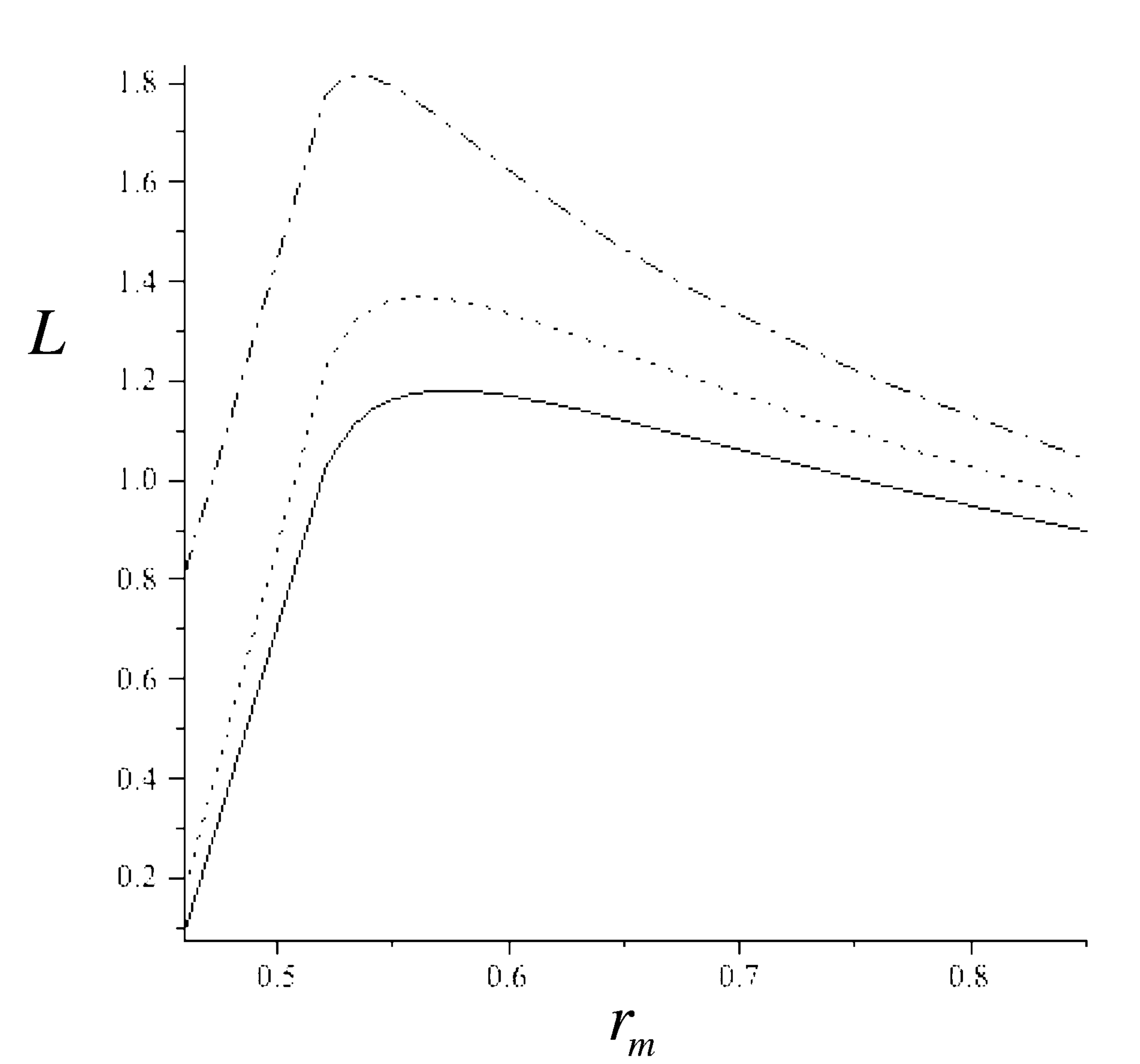}
	\caption{Dependence of the boundary length $L$ between two particles as a function
	of the string midpoint $r_m$  for $k=1$ (solid), $k=0$ (dot), and $k=-1$ (dot-dash), where $r_0=0.5$. These curves rapidly become indistinguishable as $r_0$ increases.}
	\label{f8}
\end{figure}
\begin{figure}
	\includegraphics[scale=0.15]{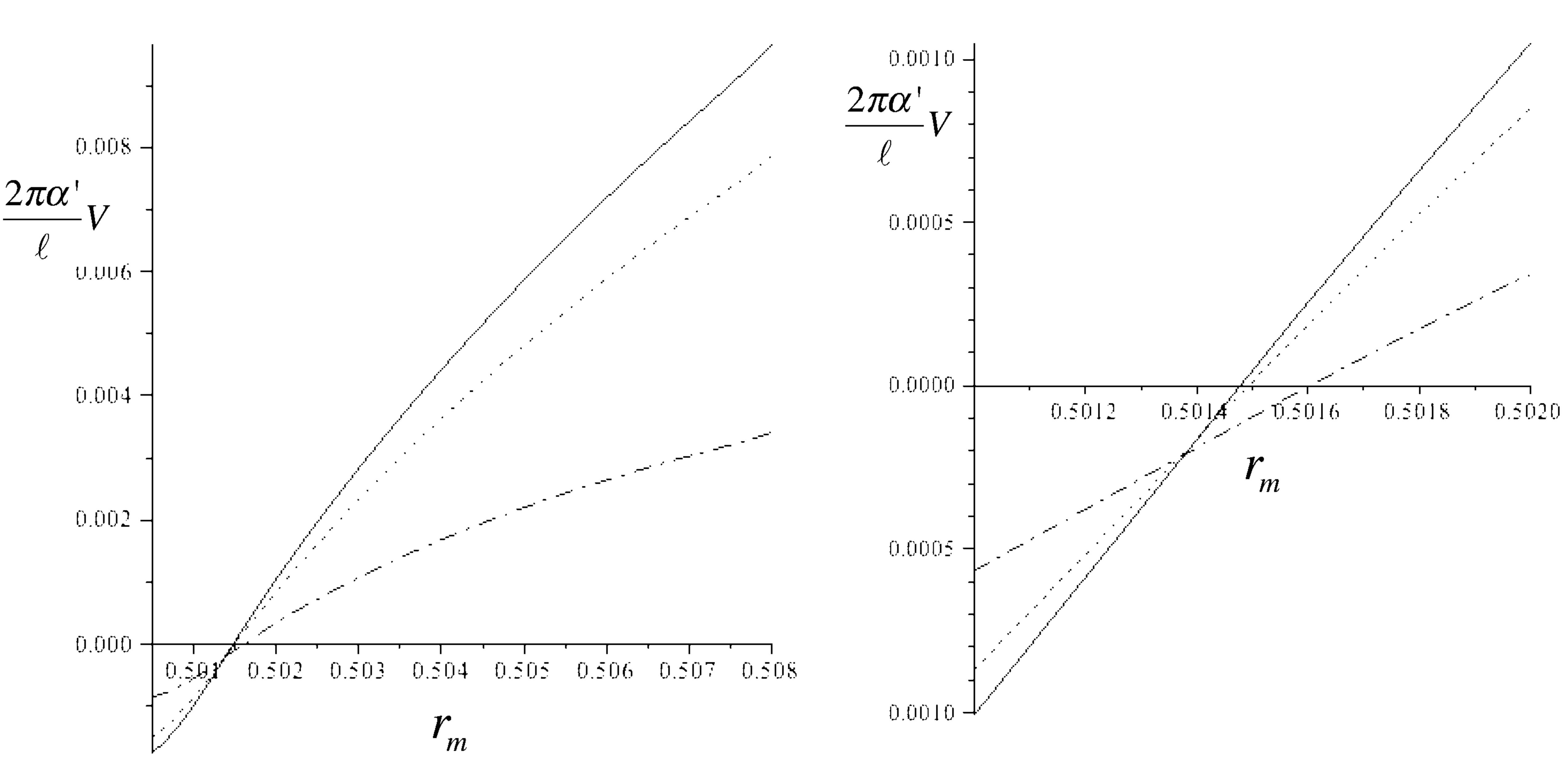}
	\caption{Dependence of the potential $V$ between two particles as a function
	of the string midpoint $r_m$  for $k=1$ (solid), $k=0$ (dot), and $k=-1$ (dot-dash), where $r_0=0.5$.
	The right-hand diagram is a close-up of the left.  The critical values of $r_m$ for each $k$ are within 
	0.04\% of each other.  The intersection points (and the behaviour of $V$) rapidly become indistinguishable as $r_0$ increases. }
	\label{f9}
\end{figure}

For the exact solution (\ref{eq20}) it is straightforward to carry out the integration in equations (\ref{eq28}) and (\ref{eq29}), though the results cannot be obtained as an explicit function of $r_m$ but must instead be obtained numerically.   It is straightforward to check that the integrands (and hence the integrals) in equations (\ref{eq28}) and (\ref{eq29}) are identical within limits of tolerance for the exact solution  (\ref{eq20})  and its
numerical counterpart with $r_0=1/\sqrt{2}$.   The behaviour is not too different for this
case as compared to its $k=0,1$ counterparts, and so the results are not plotted here.

\section{Conclusions}

In this paper I have expanded the class of  black holes that provide a dual description of a  finite temperature a Lifshitz system to include black holes of any topology.   The gravitational theory can at least be
regarded as a
phenomenological description of the 2+1 dimensional physics described by the Lifshitz theory.  Whether or not this duality can be fully incorporated into string theory, thereby extending the AdS/CFT correspondence, is an open question.

Thermodynamic properties of these black holes are quite similar for large black holes, but differ considerably for small black holes.   The genus 0 and 1 cases approach extremality as the black hole size approaches zero. However the higher-genus black holes approach extremality at $r_0=1/\sqrt{5}$, at which point the gauge fields vanish and a negative mass topological AdS black hole is attained.

The screening behaviour of the dual theory is essentially the same for any genus for large black holes.  For small black holes the onset of screening is attained at the nearly same value of the midpoint $r_m$ of the string joining two charged particles on the boundary regardless of the genus;  however this yields very different values of the critical length $L$ between the particles on the boundary due to a sensitive dependence of this quantity on $r_m$.  

A number of interesting questions remain, including higher-dimensional generalizations, a more complete study of the $z\neq 2$ cases, developing holographic renormalization for this class of theories, and understanding better the relationship with non-relativistic,
non-abelian gauge theories having quantum critical behavior at $z=2$ \cite{nonz2}.
 
\section*{Acknowledgements}
I am grateful for the hospitality of the Kavli Institute for Theoretical Physics where this work was carried out, and to the Fulbright Foundation and the Natural Sciences and Engineering Research Council of Canada for financial support.

\bibliographystyle{plain}

\end{document}